# Critical Success Factors for Positive User Experience in Hotel Websites: Applying Herzberg's Two Factor Theory for User Experience Modeling

Arunasalam Sambhanthan*
School of Computing, University of Portsmouth, UK

Alice Good
School of Computing, University of Portsmouth, UK

**ABSTRACT**

This research presents the development of a critical success factor matrix for increasing positive user experience of hotel websites based upon user ratings. Firstly, a number of critical success factors for web usability have been identified through the initial literature review. Secondly, hotel websites were surveyed in terms of critical success factors identified through the literature review. Thirdly, Herzberg's motivation theory has been applied to the user rating and the critical success factors were categorized into two areas. Finally, the critical success factor matrix has been developed using the two main sets of data.

Keywords: User Experience, Herzberg's Theory, Critical Success Factors, Usability Evaluation, e-Commerce

1. **Introduction**

Website usability plays a major role in building and nurturing an effective electronic business/customer relationship. In particular, the ability to trigger a positive user experience is a vital requirement for any e-commerce websites to reach its critical mass. There are however documented issues with the usability of hotel websites (Sambhanthan *et al*, 2012; iPerceptions Inc, 2012. & Ip, Law & Lee, 2010). The ability to trigger a positive user experience via an e-commerce interface is critical in motivating the customer to buy tourism products. In other words, the user's motivation to buy is highly dependent upon how positive their experience is with the user interface. Consequently, developing web interfaces in a manner which could trigger positive user experience is critical for tourism businesses. Therefore the research questions of this study:

- What are the Critical Success Factors for Positive User Experience in Hotel websites?
- How the Herzberg's two factor theory could be applied for User Experience Modeling?

Hence, this study aims to explore the critical success factors for positive user experience in hotel websites. The study, further investigate on classifying the critical success factors according to Herzberg's two factor theory of motivation. The main contribution of this paper is to provide a classification of critical success factors into two types according to the two factor theory propounded by Herzberg.

The paper begins with the introduction and moves into the core theories. The review of existing literature in the area is placed in the next section. The hypothesis were developed next to that and followed by the methodology section. Then the results are presented and followed by statistical analysis and Herzberg's theory has been applied and the factors are classified accordingly. Finally the implications are described and followed by the conclusions and future research directions.

## 2. A Contextual Definition of Usability

Preece *et al.*, (2002) highlights that "Ensuring that interactive products are easy to learn, effective to use, and enjoyable from the user's perspective" The research presented within this paper focuses on hotel web interfaces and specifically looks at the user's experience triggered by the hotels' websites. The importance of matching usability with user experience goals is well documented (Schneiderman & Plaisant, 2005, & Preece *et al*, 2002). Preece *et al.*(2002) highlights six usability goals and ten user experience goals as the main parameters of measuring usability as depicted in the figure 1.

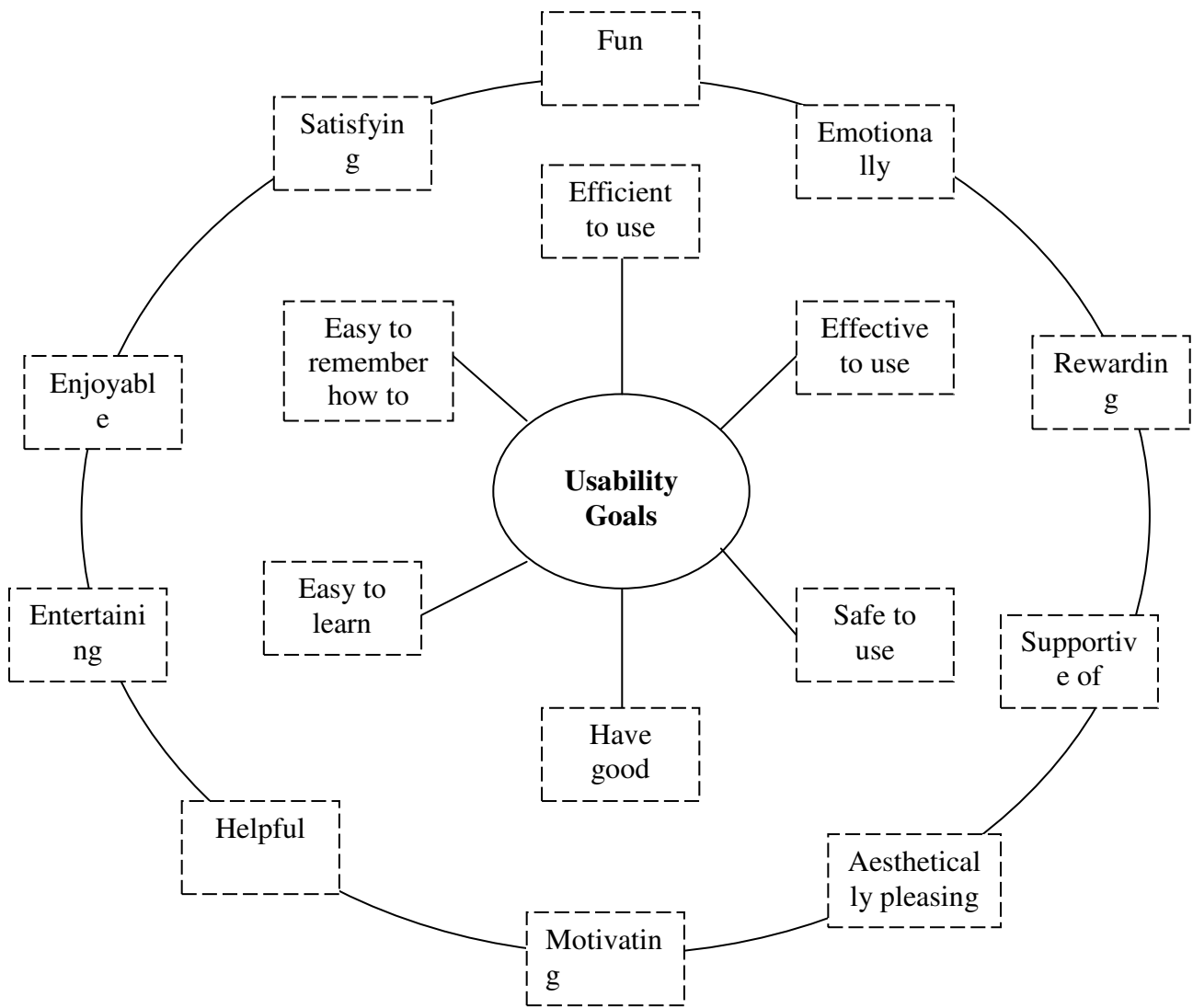

Figure 1: Usability Goals: at Center of Interaction Design
User-experience Goals: Outer Ring of Diagram (Secondary to Usability Goals)

A positive user experience is an important factor to ensure the continued usage of any interactive system. This is particularly relevant for ecommerce sites. If the site does not induce a positive user experience from the outset, customers are far less likely to return. Poor user experience of ecommerce sites can potentially contribute to loss of revenue. For the purpose of this research, the degree of usability and user experience will relate to *the specific features of hotel web sites, which trigger positive user experience through ensuring the interface is easy to learn, effective to use and enjoyable from the user's perspective*. This research aims to specify the usability shortcomings of a study of hotel websites through the identification of design elements which trigger positive user experience. The main focus of this study will therefore be on user perceptions of web interface elements affecting usability and user experience from browsing Sri Lankan hotel web sites. The results of the study can be used to inform the design of hotel websites to ensure better usability and user experience. The ten user experience goals depicted in the above diagram are taken forward for the evaluation.

### 3. Usability Evaluation Methods

Methodologies that evaluate usability and user experience are varied and tend to be user-centered by definition. Usability evaluation methods were developed in conjunction with the birth of human computer interaction. The focus was traditionally upon lab-based usability evaluation, as documented in Card *et al* (1983) and Carroll *et al* (1992). Other early methods of usability evaluation include the thinking aloud protocol (Nielsen & Molich, 1990) and shortly after, Nielsen's widely used heuristics (Nielsen, 1992).

Many methods tend to be applicable to the usability evaluation of most interfaces. However there are a few which have been designed exclusively for tourism sites. Since it commenced around 1995, the growth in tourism marketing via the Internet, led quickly to the development of evaluation methodologies, soon after in 1996. This led to pioneer work by Murphy *et al* which sought to evaluate early web development in tourism and hospitality sites (Murphy *et al*, 1996). Since then, there has been a number of evaluation frameworks developed specifically for e-commerce websites. These include a variety of methods such as surveys, case studies, observation studies, evaluation frameworks and customer satisfaction studies. Lu *et al* (2002) classified ecommerce Website evaluation into four main areas: application functionality evaluation; cost benefit analysis; user satisfaction assessment and success factors identification, whereas other researchers looked at network statistics (Fletcher et al, 2002). Later research looks at recognizing the importance of correctly defining user requirements to ensure a good user experience and usability (Preece et Al, 2007). However, research into tourism Website evaluation is still limited and Law et al (2010) suggest in their paper that specific standards for tourism Website evaluation would be useful. These standards should be interdisciplinary in their approach and essentially be human centered.

In relation to the usability evaluation of hotel sites, there has been research which emphasizes three criteria namely: (1) user interface, (2) information quality and, (3) service quality (Essawy, 2006). However, the above classification is very broad and shows very little operational focus. Additionally, the focus of this research is on the evaluation of web interface design elements, which trigger positive or negative user experience. Therefore, it is essential to evaluate the available approaches to test the usability of hotel web interfaces rather than just exploring the usability frameworks itself. Consequently, the following section comparatively analyzes the usability assessment approaches available in the past literature.

Formal analysis, automated analysis, empirical analysis and heuristic analysis are the four main methods propounded by early researchers for evaluating a user interface (Nielsen and Molich, 1990). However, Law *et al*, (2010) classifies evaluation methods into five main categories namely, counting, automated, numerical computation, user judgment, and combined methods. Apart from other approaches, heuristic analysis and user centered evaluation tend to be the two dominating methods used by researchers. Both these methods have different contextual significance as one include users in the evaluation while the other one relies upon domain experts. There is however an increasing trend among researchers to adapt innovative approaches for interface evaluation through the combination or modification of above approaches.

### 3.1. Heuristic Evaluation

Heuristic is a well-developed approach to both inform and evaluate the usability of an interface. A number of domain experts will be asked to develop criterion of design aspects which are hazardous to the usability performance of interface. The interface is then analyzed based on the heuristics and the shortcomings are then identified. In most cases, heuristics tests are being done collectively to increase the efficiency of evaluation. A plethora of research has been reported on the ways and means of heuristic evaluation (Thovtrup and Nielsen, 1991), (Nielsen, 1992), (Dzida, 1996) (Allen *et al*, 2006), (Paddison and Englefield, 2004). In addition, different scholars have developed specific heuristics and different reporting Medias relevant to certain contexts, however, Hvannberg *et al.* (2007), reports there are no significant difference between different heuristics and reporting Medias in terms of effectiveness, efficiency and inter-evaluator reliability.

Another version of the heuristic method uses an adapted approach named as 'modified heuristic approach' covering both domain experts and users in the evaluation (Yeung and Law, 2006). This approach allows more flexible evaluation using not only a sample of domain experts but also potential end users. Whilst this emulates a user centered approach, the method would obviously be more time consuming and costly, than the traditional heuristic approach, due to the incorporation of empirical evaluation. Also, the heuristics given to both types of evaluators (experts & users) should be the same to ensure the effectiveness of validation protocol.

### 3.2. Heuristic and User Testing: A Complementary Approach

More recently the research has seen the development of a framework for the evaluation of web site usability, which combines both heuristic evaluation and analytic hierarchy process (Delice and Gungor, 2009). This study approaches usability issues in two different aspects. Firstly, the exact usability problem is identified through a heuristic evaluation and then the severity listing of the same is carried out by means of circulating a questionnaire among evaluators, just after the heuristic evaluation occurs. This approach however does not directly include the users in the evaluation. It could be argued that the absence of direct user input could pose challenges in assessing the user experience. Particularly given that it is well documented that the inclusion of users within the design process does contribute to web usability (Preece *et al.*, 2007, Verdenburg *et al.*, 2002 & Good, 2011). Inspection methods, however, such as heuristic evaluation or cognitive walkthrough have been proposed within the User Centered Design paradigm. They are User- Centered because they focus on an evaluation for the user. The user is not necessarily directly involved, but evaluators must have knowledge of the user profile and evaluate what they consider to be relevant to users (not themselves). Furthermore, Nielsen argues that even users, with a minimum training, could apply the method. It is also important to consider user centered design as this is essential for search engine optimization of the sites as well (Spink, 2002). User testing could then be the effective method for an existing website while, heuristic would address the new website (Tan *et al*, 2009). Early stages in website design needs expertise recommendations on what is possible – hence heuristics to inform design. Built websites require evaluation based on user perception of what is required for any improvements – hence user testing. It could then be concluded that both of the aforementioned methods are complementary, instead of competing (Tan *et al*, 2009).

## 4. Herzberg's Two Factor Theory

Zang *et al.* (2000) argues that the presence of hygiene factors would provide the basic functionality of a website, while their absence would create user dissatisfaction. The concept of hygiene factors originates from Herzberg's motivation/hygiene theory, also known as the two factor theory (Herzberg, 1968).Motivating factors are those that contribute to user satisfaction. The study conducted by Zang et al (2000) established 44 core features in the web environment classified into 12 categories by subjects. The preliminary results show that 4 categories and 14 features were judged to be primarily motivational, while 3 categories and 13 features were perceived to be primarily hygiene in nature. The remaining 5 categories and 17 features were perceived to be both motivational and hygiene in nature. According to Zang and Darn (2000), the motivators and hygiene factors are subject to change according to the context as well as time. It should be noted that the above study had been conducted based on CNN.com, therefore a news media site. However, applying the above thinking to hotel domain is an open area for exploration. The Herzberg's two factor theory could be successfully applied to classify the criticality of factors affecting user experience in a hotel website. In particular, the competing factors of a hotel website could effectively be classified into two categories by adapting the above approach.

## 5. Critical Success Factors for Web Usability

An early definition for Critical Success Factors (CSF) suggests that there are a few key areas of activity in which favorable results are absolutely necessary for a particular manager to reach his goals (Bullen and Rockart, 1981). In the context of this research, CSF means *the factors which determine user's experience when interacting with the web interfaces of hotels*. Furthermore, the positive user experience is key for user motivation to buy. In other words, the user experience directly influences the purchase decision of the user. Hence, CSF in this research is the factors which determine positive user experience at the web interface of hotels. This research specifically focuses on developing a CSF metric based on Herzberg's two factor theory of motivation. The paper will later highlight how specific factors identified from the literature are hypothesized to check their impact on triggering positive user experience.

### 5.1. Click Stream Paradox and Security

(Nielsen, 2008) argues that the sites which take more than five clicks to reach any specific information are not usable. Furthermore, Essawy (2006) contextualized the common web usability concepts to hotel industry, arguing sites take more than three clicks to reach the desired information will be discarded by the consumers. It can however be contradictory to balance the number of clicks with the degree of security in authentication points. In fact, much information needs to be verified in different levels to ensure proper authentication. Eventually, the number of gateways passed in making a reservation will intrinsically build a confident and secure perception in the consumer psyche. It is a given that spending more time on browsing basic information will frustrate users and increase the likelihood of users switching to another site. Therefore, the basic information should be presented adhering to the specifications of Essawy, (2006) but not the reservation portals. It is also documented that a usage-oriented hierarchy or a combined hierarchy is a navigation structure associated with significantly higher usability than subject-oriented hierarchies (Fang and Holsapple, 2007). Therefore, the number of clicks taken to reach a specific set of basic information is a critical success factor. In addition to this, the secure perception is another critical success factor, which could be influenced by the number of clicks taken to make actual purchase with credit cards.

### 5.2. Value and Information Accuracy

Prior research states that the web content should be regularly updated, informative and personalized in a manner, which could directly influence the customer perceived image of destinations to create a positive virtual experience (Kozak *et al*, 2005 and Doolin *et al*, 2002). However, the updated information should be valuable to the user to make purchase decisions. Sites with irrelevant information or over informative sites could trigger negative user experience. In addition to this, information accuracy plays a vital role in ensuring usability of sites. Potentially, inaccurate information could mislead customers and lead to issues such as incorrect navigation or incorrect product selection. The accuracy of information might contradict with value in some instances where the information needs to be presented as an invitation to treat in the marketing perspective. This contradiction needs to be resolved in ensuring positive user experience in website interfaces. Therefore, value and information accuracy are another two CSF in ensuring positive user experience in tourism websites.

### 5.3. Interactivity and Loading Speed

Doolin *et al* (2002) claims interactivity of a website as the major contributor towards the quality of service itself. In broader terms, interactivity could however mean both the interactivity of the interface as well as the interactive communication between the hotel and the user through the interface. In this research, interactivity is interpreted as the interactivity of interface. Interactivity plays a major role in building up user experience. Less interactive sites may create unpleasant browsing experiences to users. However, consideration to loading speed should be coupled with interactivity to achieve the optimum outcome. A website built with multimedia features and interactive chat facilities, but lacking to have proper loading speed will undoubtedly frustrates the user. Site interactivity is then proposed as the next CSF for ensuring positive user experience.

### 5.4. Purchase Influence and Recommend-ability

Consumer ratings on the site's purchase influencing ability could be utilized to measure whether the interface has achieved its ultimate aim. Hence, the ultimate aim of web interfaces on hotels is to influence the purchase decision of browsers through improving the ease of use ability. Therefore, the ratings on sites' purchase influencing ability could be utilized as an overall assessment for the achievement of ultimate goal of site through user centered design. Yaobin (2007) in fact reports a correlation between the purchase intention and perceived ease of use of commercial websites. Purchase influencing ability could therefore be considered as another CSF.

## 5.5. Control Variables

Interestingly, Iliachenko (2006) uses, 'recommend a friend', and 'revisit intention' rating as control variables to measure the electronic service quality of tourism sites. Evidently then, the above two ratings could clearly reflect the overall consumer perception on the site. A customer will likely not revisit a site which has not provided a good user experience.. Similarly, a customer will likely not recommend the site to anyone unless they genuinely feel a positive experience when browsing the site. Therefore, both these ratings could be used as control variables for this study. However, the revisit intention could not be expected from the sample of this study consists of non-tourist audience while 'recommends a friend' rating could be considered as the browser could recommend the business to others.

## 6. Critical Success Factors

| Critical Success Factor | Description |
|---|---|
| Click Stream Paradox | The number of clicks taken for reaching a desired web location |
| Security | The degree of security involved in authentication points |
| Value | The value of information provided in the websites |
| Information Accuracy | The accuracy of information presented in the websites |
| Interactivity | The interactivity of the web interfaces |
| Loading Speed | The loading speed of the web sites |
| Purchase Influence | The ability of the websites to influence purchase decision |
| Recommend-ability | The ability of websites to trigger the user's recommendation of business to others |

Table 1: List of Critical Success Factors

As a result of the above literature findings a pilot version of the CSF matrix has been developed reflecting the CSF elements discussed in the literature review. The factors are classified into two categories as critical success factors and control variables. The matrix will be tested using a survey to develop an improved version of this matrix which will apply Herzberg's two factor theory to classify the factors depicted in blue ink.

|  | **Factors (To be categorized)** | |
|---|---|---|
| **Critical Success Factors** | Click Stream | Security |
|  | Value | Information Accuracy |
|  | Interactivity | Loading Speed |
| **Control Variables** | Purchase Influence (To be rated) | |
|  | Recommend-ability (To be rated) | |

Figure 2: Critical Success Factor Matrix – Pilot Version

## 7. Hypothesis Development

The hypotheses of this research were driven by the need to re-evaluate the criticality of the above competing factors (i.e. click stream paradox vs. security) in terms of critical success factors and hygiene factors. The critical success factors compiled in the pilot version of the matrix are synonymous with each other. Firstly, click stream paradox have a high relevance to security perception. One factor needs to be featured as dominating in order for it to be categorized as motivator and other to be labeled as hygiene factor. This applies to other two sets of factors as well. Hence, the following testable propositions were developed, which will be addressed eventually during the course of this research.

**Hypothesis 1:** *Click Stream Paradox has domination over secure perception in terms of positive user experience.*

**Hypothesis 2:** *Value has domination over Information Accuracy in terms of positive user experience.*

**Hypothesis 3:** *Interactivity has domination over loading speed in terms of positive user experience.*

**Hypothesis 4:** *Purchase Influence has domination over Recommend-ability in terms of positive user experience.*

The above four hypothesizes were divided into ten sub hypotheses, each one of those could be tested through separate ANOVA tests. The rationale for doing this is to test each of the user experience types independently in terms of the effect on critical success factors. The table presented shows the approach adapted for coding the hypothesizes. The hypothesis codes were developed by combining each hypothesis with experience types. The same codes are used throughout the data analysis phase. Refer to the appendix for a list of testable questions derived from the four propositions depicted above.

|    | E1   | E2   | E3   | E4   | E5   | E6   | E7   | E8   | E9   | E10   |
|----|------|------|------|------|------|------|------|------|------|-------|
| H1 | H 1.1 | H 1.2 | H 1.3 | H 1.4 | H 1.5 | H 1.6 | H 1.7 | H 1.8 | H 1.9 | H 1.10 |
| H2 | H 2.1 | H 2.2 | H 2.3 | H 2.4 | H 2.5 | H 2.6 | H 2.7 | H 2.8 | H 2.9 | H 2.10 |
| H3 | H 3.1 | H 3.2 | H 3.3 | H 3.4 | H 3.5 | H 3.6 | H 3.7 | H 3.8 | H 3.9 | H 3.10 |
| H4 | H 4.1 | H 4.2 | H 4.3 | H 4.4 | H 4.5 | H 4.6 | H 4.7 | H 4.8 | H 4.9 | H 4.10 |

Table 2: Hypothesis Matrix

## 8. Methodology

The aim of this study is to establish the critical success factors of hotel websites in a user perspective. Initially a comprehensive literature review has been conducted and a set of hypothesis were developed as the results of the literature review. The hypothesis were formulated based on the six critical success factors developed as the result of literature review. The user experiences triggered by specific web content criteria were measured via a web survey. A user-centric approach was adopted for this study, involving real users. A web survey was selected as the suitable method for data collection due to the fact that it could be used to collect data from geographically dispersed users. Data gathered through the web survey includes the web content rating and user experience ratings of users. The web content rating was collected to measure the effectiveness of web content in ensuring usability of the site. The user experience ratings were collected to measure the kind of experience triggered by specific web content criteria.

**8.1. Research Instrument**

The instrument for this research is a web based survey. The survey consists of three main sections. The first section was designed to track the participant profile. The second section focused on the user ratings of web content. The third section focused on user experience ratings.

Likert questions were designed in a five point scale from strongly agree to strongly disagree with a neutral scale falling on three, for tracing the user ratings on web content. The designed questionnaire was piloted with a student, IT professional, housewife, senior citizen, lecturer, statistician and a usability scholar of the university to ensure its usability for users with varied IT literacy backgrounds, analytical feasibility of the questions and its interface usability. Improvements were made according to the comments given by the above. A website was created to educate the respondents about the survey. The site was specifically designed to receive the informed consent of participants for the questionnaire study.

One hundred and forty three participants took part in this survey. While there may be some criticism of selecting a sample of generic users instead of tourists, it is defendable in this instance. The purpose of study is to measure the user experience triggered by the sites. User experience of a site is largely depending on design aspects but not on whether the user is a tourist or non-tourist. In fact, the usability of a site would be same for a tourist and non tourist user. However, the sample includes 10 users having membership in virtual tour communities. This is to ensure the sample includes a number of users having considerable interest on tourism. The geographical area of the study included Australia, Canada, Ghana, India, Malaysia, Nepal, Norway, Saudi Arabia, Seychelles, Sri Lanka, Sweden, UAE, UK and USA. (Refer figure 3). The participant profile is presented in the table 2.

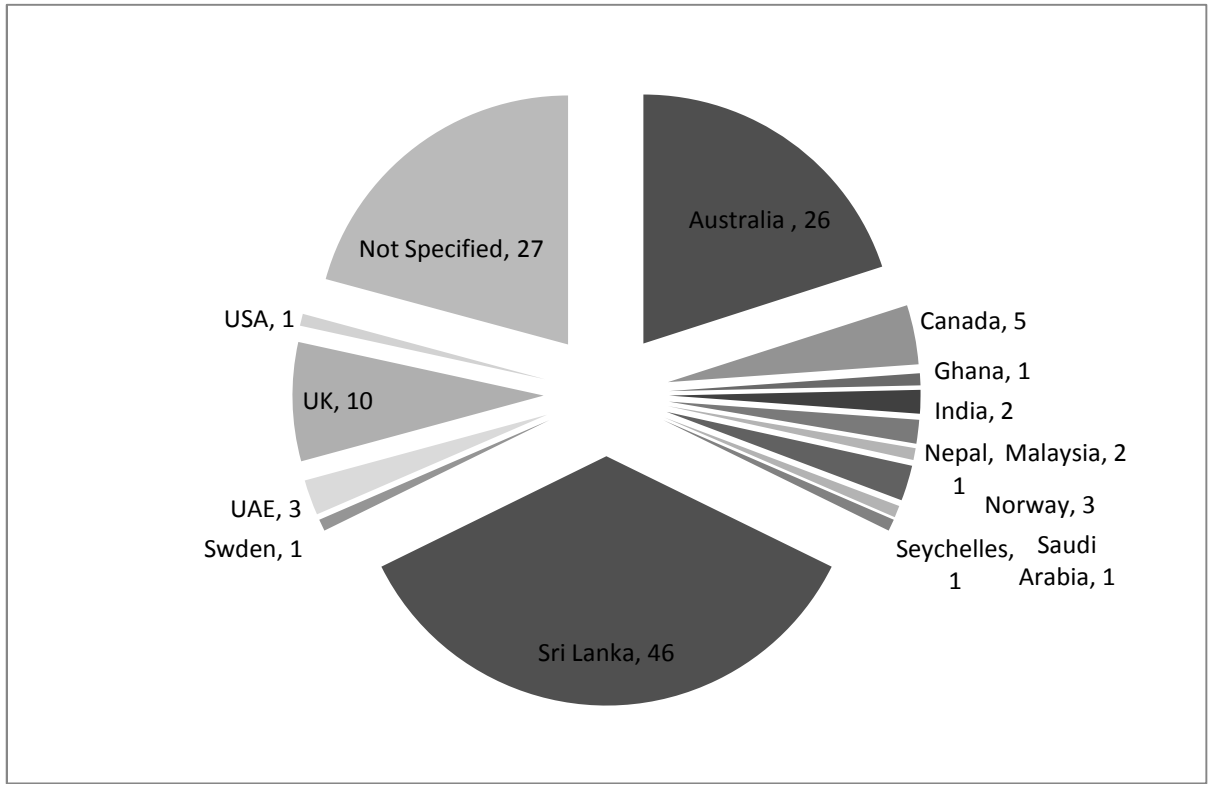

Figure 3: Country wise Respondent Profiles

| Measure | Value | Frequency | Percentage |
|---|---|---:|---:|
| Country of Residence | Sri Lanka | 46 | 35.38% |
|  | Foreign Country | 84 | 81.55% |
|  |  |  |  |
| Type of connection | Dial Up | 3 | 2.80% |
|  | ADSL | 49 | 45.79% |
|  | Broad band | 54 | 50.47% |
|  | Any other | 1 | 0.93% |
|  |  |  |  |
| Average time spent on Browsing | < 30 minutes | 7 | 6.67% |
|  | 30 minutes – 1hr | 18 | 17.14% |
|  | 1 hr - 2 hr | 30 | 28.57% |
|  | 2 hr - 3 hr | 21 | 20.00% |
|  | > 3 hr | 29 | 27.62% |
|  |  |  |  |
| Number of Search Results checking | Up to 10 | 60 | 57.69% |
|  | Up to 20 | 26 | 25.00% |
|  | Up to 30 | 9 | 8.65% |
|  | Up to 40 | 5 | 4.81% |
|  | Up to 50 | 4 | 3.85% |
|  |  |  |  |
| Search Concern | Time to load | 14 | 13.46% |
|  | Innovative Design | 5 | 4.81% |
|  | Reliable Information | 68 | 65.38% |
|  | Virtual Experience | 5 | 4.81% |
|  | Any other | 12 | 11.54% |
| Virtual Community Membership | Member | 11 | 10.48% |
|  | Non Member | 94 | 89.52% |
| Special Needs | Yes | 17 | 16.35% |
|  | No | 87 | 83.65% |
| Special Needs type | Visual | 6 | 35.29% |
|  | Auditory | 1 | 5.88% |
|  | Mobility | 5 | 29.41% |
|  | Cognitive | 3 | 17.65% |
|  | Any other | 2 | 11.76% |

Table 3: Survey Participant Characteristics

## 8.2 Selection of Samples

The list of Sri Lankan hotels was obtained from the Sri Lanka Tourist Board (SLTB) and western region hotels were abstracted from the list. A Google search was done for all hotels and the hotels holding an e-commerce site were short listed from the abstracted list. The number of questionnaires to be promoted was processed as 120. In fact the maximum number of dependent variables was determined as 4. Altogether 80 responses were estimated deciding 20 responses per variable. To reach 80 responses it was decided to gather 120 considering 40 unusable questionnaires. Also, 10 user samples were planned to represent the tourist community. Considering the above outcome the survey was planned for one full week. Links of all 5 hotel sites were given to the respondents and they were asked to select one site to record their browsing experience.

## 8.3. Two Factor Theory Applications

The CSF was tested using ANOVA and the list of CSF was separated into two categories by applying Herzberg's two factor theory. The categories are classified into two as motivators and hygiene factors. Herzberg's theory of motivation has been applied at this point for the classification of factors. The factors were classified based on the p values derived in the ANOVA test. The influence of one factor over the other one has been determined based on looking at whether the p value is > 0.5 or < 0.5. The hypothesis classification has been structured into two categories based on the aforementioned classification.

## 9. Results

The descriptive statistics derived from the data is presented in the appendix. The statistical mean and median values have been derived from the data and presented in tables 3, 4, 5 and 6. The descriptive statistics of positively influenced users, (depicted in tables 3 and 4) shows that almost all of the mean values are above average [3.00]. Although the same pattern is observed with negatively influenced users (depicted in tables 5 and 6), there are some exceptions observed in *security, purchase influence* and *recommend-ability*.

Almost all medians are on average for purchase influence rating. Majority of the means values, except emotional fulfillment and fun show below average. Therefore, the sites' ability to influence the purchase decision is identified as poor. All medians show average value for security rating. But, all experience categories except emotional fulfillment, fun and entertainment, shows a below average mean value. This shows a negative user perception on security, regardless of the above average ratings for three web content criteria. In fact, it is generalized as there is a significant correlation between users' security perception and user experience. Results depicts, users with negative experience have rated negatively for 'Recommend' criteria as well. Although the median values of the above criteria shows an average rating, the mean value records for a poor rating by the negatively experienced users. Therefore, the recommend-ability of the site has been scrutinized as a shortcoming which triggers negative user experience.

| User Experience | Value | | Accuracy | | Interactivity | | Loading Speed | |
|---|---|---|---|---|---|---|---|---|
| | Mean | Median | Mean | Median | Mean | Median | Mean | Median |
| **Satisfaction** | 3.67 | 4.00 | 3.46 | 3.00 | 3.48 | 4.00 | 3.21 | 3.00 |
| **Enjoy** | 3.69 | 4.00 | 3.48 | 3.00 | 3.49 | 4.00 | 3.25 | 3.00 |
| **Fun** | 3.66 | 4.00 | 3.47 | 3.00 | 3.50 | 4.00 | 3.27 | 3.00 |
| **Entertainment** | 3.68 | 4.00 | 3.48 | 3.00 | 3.51 | 4.00 | 3.29 | 3.00 |
| **Independence** | 3.66 | 4.00 | 3.44 | 3.00 | 3.47 | 4.00 | 3.18 | 3.00 |
| **Motivation** | 3.67 | 4.00 | 3.46 | 3.00 | 3.48 | 4.00 | 3.21 | 3.00 |
| **Aesthetically Pleasing** | 3.68 | 4.00 | 3.48 | 3.00 | 3.51 | 4.00 | 3.29 | 3.00 |
| **Rewarding** | 3.68 | 4.00 | 3.48 | 3.00 | 3.51 | 4.00 | 3.29 | 3.00 |
| **Emotionally Fulfilling** | 3.69 | 4.00 | 3.47 | 3.00 | 3.50 | 4.00 | 3.31 | 3.00 |
| **Revisit** | 1.95 | 4.00 | 3.48 | 3.00 | 3.49 | 4.00 | 3.25 | 3.00 |

Table 4: Web Content Ratings of Users with Positive Experience

| User Experience | Click Stream | | Security | | Purchase In. | | Recommend | |
|---|---|---|---|---|---|---|---|---|
| | Mean | Median | Mean | Median | Mean | Median | Mean | Median |
| **Satisfaction** | 3.19 | 3.00 | 3.12 | 3.00 | 3.06 | 3.00 | 3.19 | 3.00 |
| **Enjoy** | 3.23 | 3.00 | 3.15 | 3.00 | 3.09 | 3.00 | 3.22 | 3.00 |
| **Fun** | 3.26 | 3.00 | 3.16 | 3.00 | 3.11 | 3.00 | 3.24 | 3.00 |
| **Entertainment** | 3.27 | 3.00 | 3.19 | 3.00 | 3.38 | 3.00 | 3.25 | 3.00 |
| **Independence** | 3.18 | 3.00 | 3.09 | 3.00 | 3.04 | 3.00 | 3.18 | 3.00 |

| User Experience | | | | | | | | |
|---|---|---|---|---|---|---|---|---|
| Motivation | 3.19 | 3.00 | 3.12 | 3.00 | 3.06 | 3.00 | 3.19 | 3.00 |
| Aesthetically Pleasing | 3.27 | 3.00 | 3.19 | 3.00 | 3.13 | 3.00 | 3.25 | 3.00 |
| Rewarding | 3.27 | 3.00 | 3.19 | 3.00 | 3.13 | 3.00 | 3.25 | 3.00 |
| Emotionally Fulfilling | 3.27 | 3.00 | 3.21 | 3.00 | 3.15 | 3.00 | 3.26 | 3.00 |
| Revisit | 3.23 | 3.00 | 3.15 | 3.00 | 3.08 | 3.00 | 3.22 | 3.00 |

Table 5: Web Content Ratings of Users with Positive Experience

| User Experience | Value | | Accuracy | | Interactivity | | Loading Speed | |
|---|---|---|---|---|---|---|---|---|
| | Mean | Median | Mean | Median | Mean | Median | Mean | Median |
| Satisfaction | 3.54 | 3.00 | 3.30 | 3.00 | 3.31 | 3.00 | 3.00 | 3.00 |
| Enjoy | 3.55 | 3.00 | 3.33 | 3.00 | 3.35 | 3.00 | 3.03 | 3.00 |
| Fun | 3.64 | 4.00 | 3.42 | 3.00 | 3.45 | 4.00 | 3.17 | 3.00 |
| Entertainment | 3.61 | 4.00 | 3.40 | 3.00 | 3.43 | 3.00 | 3.13 | 3.00 |
| Independence | 3.53 | 3.00 | 3.29 | 3.00 | 3.33 | 3.00 | 2.98 | 3.00 |
| Motivation | 3.56 | 4.00 | 3.33 | 3.00 | 3.36 | 3.00 | 3.05 | 3.00 |
| Aesthetically Pleasing | 3.58 | 4.00 | 3.37 | 3.00 | 3.40 | 3.00 | 3.11 | 3.00 |
| Rewarding | 3.55 | 3.50 | 3.33 | 3.00 | 3.35 | 3.00 | 3.03 | 3.00 |
| Emotionally Fulfilling | 3.62 | 4.00 | 3.41 | 3.00 | 3.44 | 3.50 | 3.16 | 3.00 |
| Revisit | 3.58 | 4.00 | 3.37 | 3.00 | 3.40 | 3.00 | 3.11 | 3.00 |

Table 6: Web Content Ratings of Users with Negative Experience

| User Experience | Click Stream | | Security | | Purchase In. | | Recommend | |
|---|---|---|---|---|---|---|---|---|
| | Mean | Median | Mean | Median | Mean | Median | Mean | Median |
| Satisfaction | 3.02 | 3.00 | 2.83 | 3.00 | 2.74 | 3.00 | 2.98 | 3.00 |
| Enjoy | 3.02 | 3.00 | 2.90 | 3.00 | 2.87 | 3.00 | 3.05 | 3.00 |
| Fun | 3.16 | 3.00 | 3.06 | 3.00 | 3.01 | 3.00 | 3.17 | 3.00 |
| Entertainment | 3.12 | 3.00 | 3.01 | 3.00 | 2.99 | 3.00 | 3.15 | 3.00 |
| Independence | 3.02 | 3.00 | 2.84 | 3.00 | 2.76 | 3.00 | 2.98 | 3.00 |
| Motivation | 3.03 | 3.00 | 2.92 | 3.00 | 2.89 | 3.00 | 3.05 | 3.00 |
| Aesthetically Pleasing | 3.09 | 3.00 | 2.98 | 3.00 | 2.88 | 3.00 | 3.11 | 3.00 |
| Rewarding | 3.02 | 3.00 | 2.90 | 3.00 | 2.87 | 3.00 | 3.05 | 3.00 |
| Emotionally Fulfilling | 3.15 | 3.00 | 3.03 | 3.00 | 3.00 | 3.00 | 3.16 | 3.00 |
| Revisit | 3.09 | 3.00 | 2.98 | 3.00 | 2.95 | 3.00 | 2.95 | 3.00 |

Table 7: Web Content Ratings of Users with Negative Experience

## 10. Data Analysis

The results obtained through web survey were subjected to statistical analysis. The statistical ANOVA has been conducted for each sub hypothesis and presented. The p values of each sub hypothesis are presented in the below table. The p values for each hypothesis are presented in separate graphs and followed by discussion.

## 10.1. Statistical Analysis of Hypothesis 1

| Hypothesis | Source of Variation | Sum of Squares | d.f. | Mean Squares | F | p value |
|---|---|---|---|---|---|---|
| H 1.1 | between | 0.2273 | 1 | 0.2273 | 0.534 | 0.47 |
| | error | 45.96 | 108 | 0.4256 | | |
| | total | 46.19 | 109 | | | |
| H 1.2 | between | 0.1481 | 1 | 0.1481 | 0.3876 | 0.53 |
| | error | 40.52 | 106 | 0.3823 | | |
| | total | 40.67 | 107 | | | |
| H 1.3 | between | 0.3571 | 1 | 0.3571 | 0.9551 | 0.33 |
| | error | 25.43 | 68 | 0.3739 | | |
| | total | 25.79 | 69 | | | |
| H 1.4 | between | 0.2051 | 1 | 0.2051 | 0.5084 | 0.48 |
| | error | 30.67 | 76 | 0.4035 | | |
| | total | 30.87 | 77 | | | |
| H 1.5 | between | 0.4455 | 1 | 0.4455 | 1.008 | 0.32 |
| | error | 47.75 | 108 | 0.4421 | | |
| | total | 48.19 | 109 | | | |
| H 1.6 | between | 0.1071 | 1 | 0.1071 | 0.254 | 0.62 |
| | error | 34.6 | 82 | 0.4219 | | |
| | total | 34.7 | 83 | | | |
| H 1.7 | between | 0.1023 | 1 | 0.1023 | 0.2674 | 0.61 |
| | error | 32.89 | 86 | 0.3824 | | |
| | total | 32.99 | 87 | | | |
| H 1.8 | between | 0 | 1 | 0 | 0 | 1 |
| | error | 31.95 | 82 | 0.3897 | | |
| | total | 31.95 | 83 | | | |
| H 1.9 | between | 0 | 1 | 0 | 0 | 1 |
| | error | 20.29 | 68 | 0.2983 | | |
| | total | 20.29 | 69 | | | |
| H 1.10 | between | 0 | 1 | 0 | 0 | 1 |
| | error | 31.85 | 76 | 0.419 | | |
| | total | 31.85 | 77 | | | |

Table 8: Statistical Analysis of Hypothesis 1

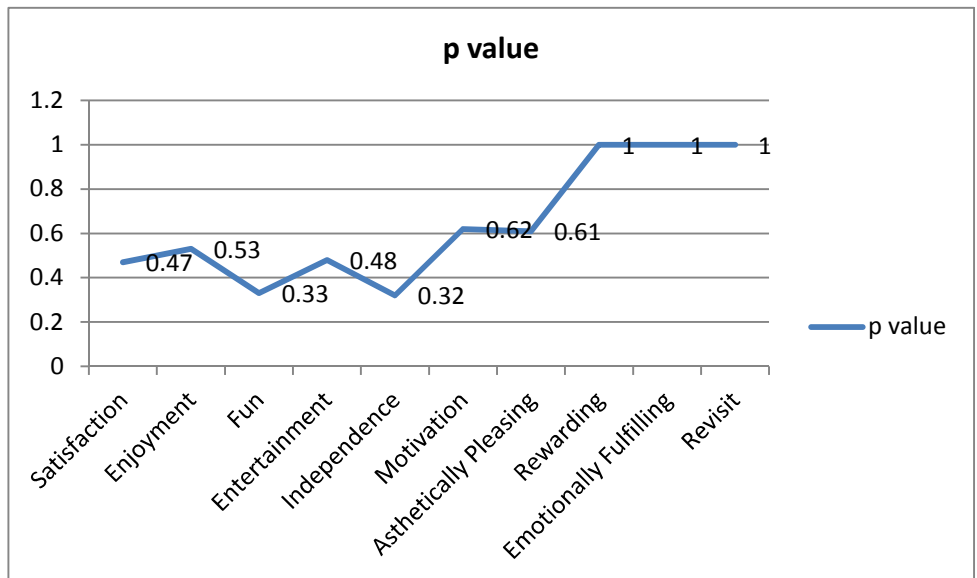

Figure 4: P values of hypothesis 1

**Hypothesis 1:** The results of six experience types shows a p value > 0.5 and four experience types shows p value < 0.5 for hypothesis 1. Therefore the majority of the sub hypotheses are valid. Therefore the research hypothesis could be accepted. Thus, it could be concluded that the click stream paradox does have a significant domination over the secure perception of users.

### 10.2. Statistical Analysis of Hypothesis 2

| Hypothesis | Source of Variation | Sum of Squares | d.f. | Mean Squares | F | p value |
|---|---|---|---|---|---|---|
| H 2.1 | between | 0.9259 | 1 | 0.9259 | | |
| | error | 39.74 | 106 | 0.3749 | 2.47 | 0.12 |
| | total | 40.67 | 107 | | | |
| H 2.2 | between | 0.9259 | 1 | 0.9259 | | |
| | error | 39.74 | 106 | 0.3749 | 2.47 | 0.12 |
| | total | 40.67 | 107 | | | |
| H 2.3 | between | 0.9143 | 1 | 0.9143 | | |
| | error | 22.57 | 68 | 0.3319 | 2.754 | 0.1 |
| | total | 23.49 | 69 | | | |
| H 2.4 | between | 1.551 | 1 | 1.551 | | |
| | error | 25.9 | 76 | 0.3408 | 4.552 | 0.036 |
| | total | 27.45 | 77 | | | |
| H 2.5 | between | 1.536 | 1 | 1.536 | | |
| | error | 41.64 | 108 | 0.3855 | 3.985 | 0.048 |
| | total | 43.17 | 109 | | | |
| H 2.6 | between | 1.44 | 1 | 1.44 | | |
| | error | 30.98 | 82 | 0.3778 | 3.813 | 0.054 |

| Hypothesis | Source of Variation | Sum of Squares | d.f. | Mean Squares | F | p value |
|---|---|---|---|---|---|---|
| | total | 32.42 | 83 | | | |
| H 2.7 | between | 1.136 | 1 | 1.136 | | |
| | error | 33.73 | 86 | 0.3922 | 2.898 | 0.092 |
| | total | 34.86 | 87 | | | |
| H 2.8 | between | 0.9643 | 1 | 0.9643 | | |
| | error | 30.6 | 82 | 0.3731 | 2.584 | 0.11 |
| | total | 31.56 | 83 | | | |
| H 2.9 | between | 1.429 | 1 | 1.429 | | |
| | error | 23.71 | 68 | 0.3487 | 4.096 | 0.047 |
| | total | 25.14 | 69 | | | |
| H 2.10 | between | 0.8205 | 1 | 0.8205 | | |
| | error | 29.64 | 76 | 0.39 | 2.104 | 0.15 |
| | total | | 77 | | | |

Table 9: Statistical Analysis of Hypothesis 2

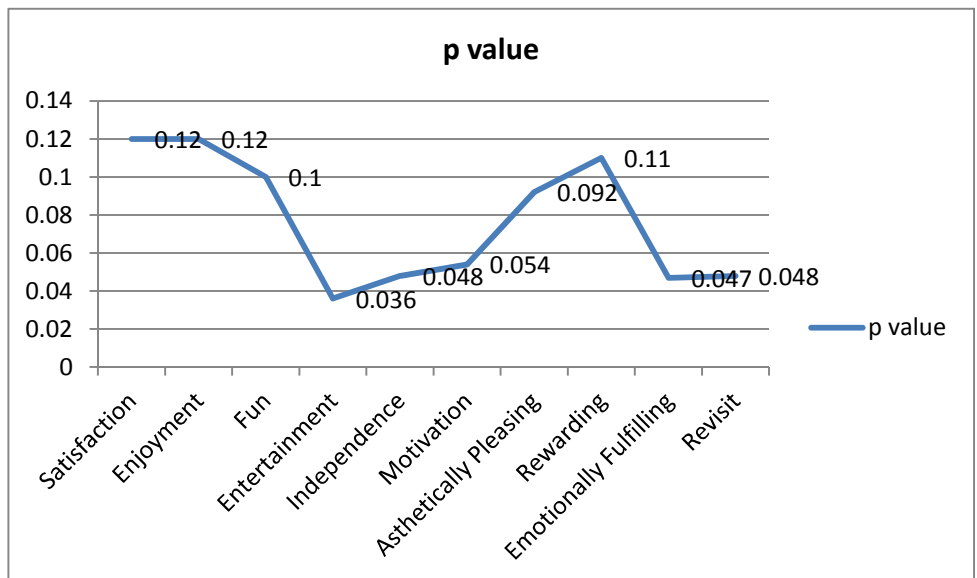

Figure 5: P values of hypothesis 2

**Hypothesis 2:** The results of all ten experience types show a p value <0.5 for hypothesis 2. Overall, this means that the probability of research hypothesis is not valid and the research hypothesis cannot be accepted. Thus, it could be concluded that value does not have a significant domination over the information accuracy.

### 10.3. Statistical Analysis of Hypothesis 3

| Hypothesis | Source of Variation | Sum of Squares | d.f. | Mean Squares | F | p value |
|---|---|---|---|---|---|---|
| | between | 1.12 | 1 | 1.12 | | |
| H 3.1 | error   | 60.31  | 106 | 0.569  | 1.969 | 0.16  |
|       | total   | 61.44  | 107 |        |       |       |
|       | between | 1.565  | 1   | 1.565  |       |       |
| H 3.2 | error   | 55.87  | 106 | 0.5271 | 2.969 | 0.088 |
|       | total   | 57.44  | 107 |        |       |       |
|       | between | 1.729  | 1   | 1.729  |       |       |
| H 3.3 | error   | 34.86  | 68  | 0.5126 | 3.372 | 0.071 |
|       | total   | 36.59  | 69  |        |       |       |
|       | between | 1.282  | 1   | 1.282  |       |       |
| H 3.4 | error   | 42.1   | 76  | 0.554  | 2.314 | 0.13  |
|       | total   | 43.38  | 77  |        |       |       |
|       | between | 1.536  | 1   | 1.536  |       |       |
| H 3.5 | error   | 67.64  | 108 | 0.6263 | 2.453 | 0.12  |
|       | total   | 69.17  | 109 |        |       |       |
|       | between | 3.44   | 1   | 3.44   |       |       |
| H 3.6 | error   | 43.26  | 82  | 0.5276 | 6.521 | 0.013 |
|       | total   | 46.7   | 83  |        |       |       |
|       | between | 2.557  | 1   | 2.557  |       |       |
| H 3.7 | error   | 41.16  | 86  | 0.4786 | 5.342 | 0.023 |
|       | total   | 43.72  | 87  |        |       |       |
|       | between | 1.44   | 1   | 1.44   |       |       |
| H 3.8 | error   | 37.45  | 82  | 0.4567 | 3.154 | 0.079 |
|       | total   | 38.89  | 83  |        |       |       |
|       | between | 0.9143 | 1   | 0.9143 |       |       |
| H 3.9 | error   | 30.23  | 68  | 0.4445 | 2.057 | 0.16  |
|       | total   | 31.14  | 69  |        |       |       |
|       | between | 2.885  | 1   | 2.885  |       |       |
| H 3.10| error   | 40.41  | 76  | 0.5317 | 5.425 | 0.023 |
|       | total   | 43.29  | 77  |        |       |       |

Table 10: Statistical Analysis of Hypothesis 3

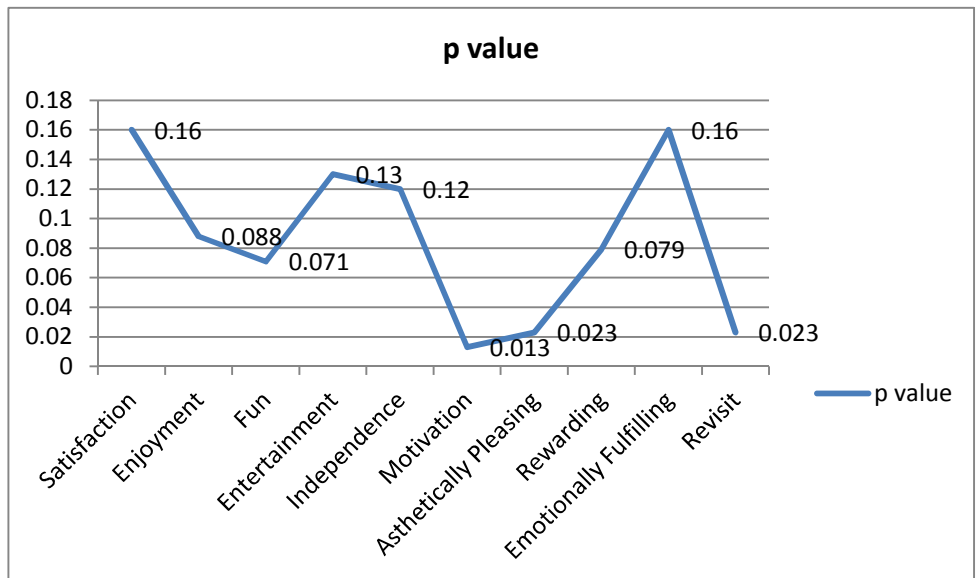
Figure 6: P values of hypothesis 3

**Hypothesis 3:** The results of all ten experience types show a p value <0.5 for hypothesis 3. This means in overall the probability of research hypothesis is not valid and the research hypothesis cannot be accepted. Thus, it could be concluded that Interactivity does not have a significant domination over the Loading Speed.

**10.4. Statistical Analysis of Hypothesis 4**

| Hypothesis | Source of Variation | Sum of Squares | d.f. | Mean Squares | F | p value |
|---|---|---|---|---|---|---|
| H 4.1 | between | 0.2315 | 1 | 0.2315 | 0.5402 | 0.46 |
|  | error | 45.43 | 106 | 0.4285 |  |  |
|  | total | 45.66 | 107 |  |  |  |
| H 4.2 | between | 0.2315 | 1 | 0.2315 | 0.5748 | 0.45 |
|  | error | 42.69 | 106 | 0.4027 |  |  |
|  | total | 42.92 | 107 |  |  |  |
| H 4.3 | between | 0.5143 | 1 | 0.5143 | 1.31 | 0.26 |
|  | error | 26.69 | 68 | 0.3924 |  |  |
|  | total | 27.2 | 69 |  |  |  |
| H 4.4 | between | 0.4615 | 1 | 0.4615 | 1.194 | 0.28 |
|  | error | 29.38 | 76 | 0.3866 |  |  |
|  | total | 29.85 | 77 |  |  |  |
| H 4.5 | between | 0.9091 | 1 | 0.9091 | 2.069 | 0.15 |
|  | error | 47.45 | 108 | 0.4394 |  |  |
|  | total | 48.36 | 109 |  |  |  |
| H 4.6 | between | 0.1905 | 1 | 0.1905 | 0.4216 | 0.52 |
|  | error | 37.05 | 82 | 0.4518 |  |  |
|  | total | 37.24 | 83 |  |  |  |
|  | between | 0.1818 | 1 | 0.1818 |  |  |
| H 4.7  | error   | 33.27  | 86 | 0.3869 | 0.4699| 0.49 |
|        | total   | 33.45  | 87 |        |       |      |
| H 4.8  | between | 0.4286 | 1  | 0.4286 |       |      |
|        | error   | 34.71  | 82 | 0.4233 | 1.012 | 0.32 |
|        | total   | 35.14  | 83 |        |       |      |
| H 4.9  | between | 0.5143 | 1  | 0.5143 |       |      |
|        | error   | 25.83  | 68 | 0.3798 | 1.354 | 0.25 |
|        | total   | 26.34  | 69 |        |       |      |
| H 4.10 | between | 1.038  | 1  | 1.038  |       |      |
|        | error   | 35.33  | 76 | 0.4649 | 2.234 | 0.14 |
|        | total   | 36.37  | 77 |        |       |      |

Table 11: Statistical Analysis of Hypothesis 4

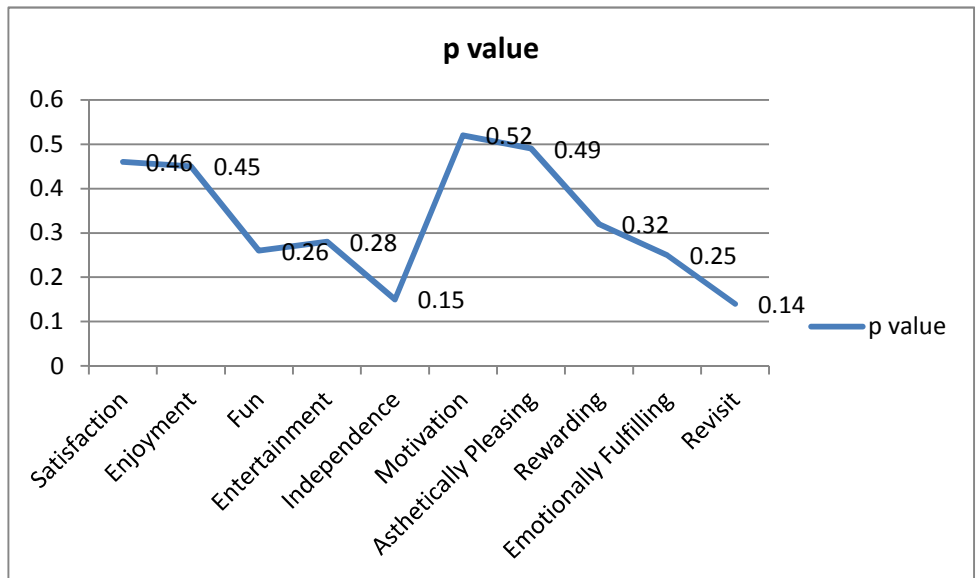

Figure 7: P values of hypothesis 4

**Hypothesis 4:** The results of all nine experience types show a p value <0.5 for hypothesis 2. This means in overall the probability of research hypothesis is not valid and the research hypothesis cannot be accepted. Thus, it could be concluded that purchase influence does not have a significant domination over the recommend - ability.

|    | E1   | E2    | E3    | E4    | E5    | E6    | E7    | E8    | E9    | E10   |
|----|------|-------|-------|-------|-------|-------|-------|-------|-------|-------|
| H1 | 0.47 | 0.53  | 0.33  | 0.48  | 0.32  | 0.62  | 0.61  | 1     | 1     | 1     |
| H2 | 0.12 | 0.12  | 0.1   | 0.036 | 0.048 | 0.054 | 0.092 | 0.11  | 0.047 | 0.048 |
| H3 | 0.16 | 0.088 | 0.071 | 0.13  | 0.12  | 0.013 | 0.023 | 0.079 | 0.16  | 0.023 |
| H4 | 0.46 | 0.45  | 0.26  | 0.28  | 0.15  | 0.52  | 0.49  | 0.32  | 0.25  | 0.14  |

Table 12: P value Summary

## 11. Model Development

The factors were organized into two groups based on the results obtained from hypothesis tests. The structural model obtained through the above exercise is presented below. Click stream paradox is selected as a motivator while the security falls under hygiene factor. Arguably, security is one of the most important concerns of users. Research by Law and Wong (2003) among Hong Kong participants shows e-buyers of travel products concern 'secure payment methods' as the most important aspect. Information Accuracy is selected as a motivator while Value falls under hygiene factor. Information accuracy in its own form is the foremost concern when it comes to reservations in terms of tourism business. Although, Value is an important factor, the accuracy remarks a distinctive place in criticality. Loading speed is selected as a motivator while interactivity gets into hygiene factor. Interactivity of the web interface could be an important factor - but loading speed is more critical for businesses to have competitive advantage over other organizations. Control Variables are used as a validation protocol to the constructs identified in the above section. Recommend-ability is a motivator while the Purchase influence falls under hygiene factor.

|  | **Motivator** | **Hygiene** |
|---|---|---|
| **Factors** | Click Stream Paradox | Security |
|  | Information Accuracy | Value |
|  | Loading Speed | Interactivity |
| **Control Variables** | Recommend-ability | Purchase Influence |

Table 13: Classified Critical Success Factors

## 12. Discussion

(Nielsen, 2008) argues that the sites which take more than five clicks to reach any specific information are not usable. Furthermore, Essawy (2006) contextualized the common web usability concepts to hotel industry, arguing sites take more than three clicks to reach the desired information will be discarded by the consumers. Accordingly, the results show that the number of clicks taken to reach a specified destination in the website is more critical for the success of website when comparing with the security. It can however be contradictory to balance the number of clicks with the degree of security in authentication points. In fact, much information needs to be verified in different levels to ensure proper authentication. Eventually, the number of gateways passed in making a reservation will intrinsically build a confident and secure perception in the consumer psyche. It is a given that spending more time on browsing basic information will frustrate users and increase the likelihood of users switching to another site. Therefore, the basic information should be presented adhering to the specifications of Essawy, (2006) but not the reservation portals. It is also documented that a usage-oriented hierarchy or a combined hierarchy is a navigation structure associated with significantly higher usability than subject-oriented hierarchies (Fang and Holsapple, 2007). Therefore, the number of clicks taken to reach a specific set of basic information is a critical success factor while the security remains as a hygiene factor. In addition to this, the secure perception is another important factor, which could be influenced by the number of clicks taken to make actual purchase with credit cards. Especially, the user needs to be empowered with a secure perception when browsing the site. Although, the companies take so many security measures; it is critical to include trust inducing design elements in the web interface, in order to maintain a secure perception in the users' mind (Sambhanthan & Good, 2012).

Prior research states that the web content should be regularly updated, informative and personalized in a manner, which could directly influence the customer perceived image of destinations to create a positive virtual experience (Kozak *et al.*, 2005 and Doolin *et al.*, 2002). However, the updated information should be valuable to the user to make purchase decisions. Sites with irrelevant information or over informative sites could trigger negative user experience. In addition to this, information accuracy plays a vital role in ensuring usability of sites. Potentially, inaccurate information could mislead customers and lead to issues such as incorrect navigation or incorrect product selection. The accuracy of information might contradict with value in some instances where the information needs to be presented as an invitation to treat in the marketing perspective. In this study, the value of information has been featured as a critical success factor while the accuracy been identified as a hygiene factor. In the context of tourism business the value and the usefulness of information is more critical compared to the accuracy of information. For an example, a customer coming to the website with the intention of room details will check for the valuable information instead of bothering whether the room rates are accurately presented. First the motivator is whether the relevant room information is presented in the web or not. But the accuracy of information is a supplementary factor which could be there as a hygiene factor.

Doolin *et al* (2002) claims interactivity of a website as the major contributor towards the quality of service itself. In broader terms, interactivity could however mean both the interactivity of the interface as well as the interactive communication between the hotel and the user through the interface. In this research, interactivity is interpreted as the interactivity of interface. Interactivity plays a major role in building up user experience. Less interactive sites may create unpleasant browsing experiences to users. However, consideration to loading speed should be coupled with interactivity to achieve the optimum outcome. A website built with multimedia features and interactive chat facilities, but lacking to have proper loading speed will undoubtedly frustrates the user. Loading speed is then proposed as the next CSF for ensuring positive user experience.

A consumer rating on the site's purchase influencing ability is utilized to measure whether the interface has achieved its ultimate aim. Hence, the ultimate aim of web interfaces on hotels is to influence the purchase decision of browsers through improving the ease of use ability. Therefore, the rating on sites' purchase influencing ability was utilized as an overall assessment for the achievement of ultimate goal of site through user centered design. However, the Recommend-ability is featured as a critical success factor while purchase influence ability is featured as a hygiene factor. Yaobin (2007) reports a correlation between the purchase intention and perceived ease of use of commercial websites. However, recommend ability is a broad concept which includes purchase influence as well as the other aspects of positivity involved in the websites. The recommend-ability could therefore be considered as another CSF.

## 13. Limitations

On evaluating the process of the above study, a number of limitations could be notified. Firstly, this study is entirely based on the ten user experience goals derived from Preece *et al.* (2001). Secondly, the target audience consists of generic users instead of tourists. Even though, the goal of study is to measure the Critical Success Factors for user experience in tourism websites, the selection of tourist samples for the study could positively influence the dependability of findings with regard to the purchase influence ratings. Thirdly, the sample size of the project consists of 120, which is adequate to make generalizations, but could be increased further to improve the accuracy of results and dependability of generalizations. Finally, the research has not been tracing a very comprehensive background variable data. It could be criticized that inferring insightful conclusions from a study which does not have much background variable is quite challenging. However, the study records this as one of the major limitations and open doors for the future researchers to rectify it in the research design process.

This research could be further developed by working on a process model. These could be derived from the results which gives a more details account of the factors presented in the structural model. Basically, the process model could be based on the relationship between each set of factors and their relationship with experience types. The process model could be used by the managers in decision making based on the contextual requirement of their hotel business. Furthermore, there are several types of tourism businesses in the hotel sector. From sport tourism to destination marketing – each type of market requires specific experience types to be generated through the website to attract the specific target groups. In such cases, there can be a need for contextual model to develop websites – since each type would require generating different user experience types through the web interfaces. Hence, the process model would be useful in developing context specific interfaces for the required categories.

# Appendix

**Hypothesis 1:** *Click Stream Paradox has domination over secure perception in terms of positive user experience.*

> **Hypothesis 1.1:** *Click Stream Paradox has domination over secure perception in terms of user satisfaction.*
>
> **Hypothesis 1.2:** *Click Stream Paradox has domination over secure perception in terms of user enjoyment.*
>
> **Hypothesis 1.3:** *Click Stream Paradox has domination over secure perception in terms of user feeling fun?*
>
> **Question 1.4:** *Click Stream Paradox has domination over secure perception in terms of user entertainment.*
>
> **Hypothesis 1.5:** *Click Stream Paradox has domination over secure perception in terms of user independence.*
>
> **Hypothesis 1.6:** *Click Stream Paradox has domination over secure perception in terms of user motivation.*
>
> **Hypothesis 1.7:** *Click Stream Paradox has domination over secure perception in terms of aesthetic pleasantness.*
>
> **Hypothesis 1.8:** *Click Stream Paradox has domination over secure perception in terms of user feeling of rewarding.*
>
> **Hypothesis 1.9:** *Click Stream Paradox has domination over secure perception in terms of user emotional fulfillment.*
>
> **Hypothesis 1.10:** *Click Stream Paradox has domination over secure perception in terms of user revisit intention.*

**Hypothesis 2:** *Value has domination over Information Accuracy in terms of positive user experience.*

> **Hypothesis 2.1:** *Value has domination over Information Accuracy in terms of user satisfaction.*
>
> **Hypothesis 2.2:** *Value has domination over Information Accuracy in terms of user enjoyment.*
>
> **Hypothesis 2.3:** *Value has domination over Information Accuracy in terms of user feeling fun.*

**Hypothesis 2.4:** *Value has domination over Information Accuracy in terms of user entertainment.*

**Hypothesis 2.5:** *Value has domination over Information Accuracy in terms of user independence.*

**Hypothesis 2.6:** *Value has domination over Information Accuracy in terms of user motivation.*

**Hypothesis 2.7:** *Value has domination over Information Accuracy in terms of user aesthetic pleasantness.*

**Hypothesis 2.8:** *Value has domination over Information Accuracy in terms of user feeling of rewarding.*

**Hypothesis 2.9:** *Value has domination over Information Accuracy in terms of user emotional fulfillment.*

**Hypothesis 2.10:** *Value has domination over Information Accuracy in terms of user revisit intention.*

**Hypothesis 3:** *Interactivity has domination over loading speed in terms of positive user experience.*

**Hypothesis 3.1:** *Interactivity has domination over loading speed option in terms of user satisfaction.*

**Hypothesis 3.2:** *Interactivity has domination over loading speed in terms of user enjoyment.*

**Hypothesis 3.3:** *Interactivity has domination over loading speed in terms of user feeling fun.*

**Hypothesis 3.4:** *Interactivity has domination over loading speed in terms of user entertainment.*

**Hypothesis 3.5:** *Interactivity has domination over loading speed in terms of user independence.*

**Hypothesis 3.6:** *Interactivity has domination over loading speed in terms of user motivation.*

**Hypothesis 3.7:** *Interactivity has domination over loading speed in terms of user aesthetic pleasantness.*

**Hypothesis 3.8:** *Interactivity has domination over loading speed in terms of user feeling of rewarding.*

**Hypothesis 3.9:** *Interactivity has domination over loading speed in terms of user emotional fulfillment.*

**Hypothesis 3.10:** *Interactivity has domination over loading speed in terms of user revisit intention.*

**Hypothesis 4:** *Purchase Influence has domination over Recommend-ability in terms of positive user experience.*

**Hypothesis 4.1:** *Purchase Influence has domination over Recommend-ability in terms of user satisfaction.*

**Hypothesis 4.2:** *Purchase Influence has domination over Recommend-ability in terms of user enjoyment.*

**Hypothesis 4.3:** *Purchase Influence has domination over Recommend-ability in terms of user feeling fun.*

**Hypothesis 4.4:** *Purchase Influence has domination over Recommend-ability in terms of user entertainment.*

**Hypothesis 4.5:** *Purchase Influence has domination over Recommend-ability in terms of user independence.*

**Hypothesis 4.6:** *Purchase Influence has domination over Recommend-ability in terms of user motivation.*

**Hypothesis 4.7:** *Purchase Influence has domination over Recommend-ability in terms of user aesthetic pleasantness.*

**Hypothesis 4.8:** *Purchase Influence has domination over Recommend-ability in terms of user feeling of rewarding.*

**Hypothesis 4.9:** *Purchase Influence has domination over Recommend-ability in terms of user emotional fulfillment.*

**Hypothesis 4.10:** *Purchase Influence has domination over Recommend-ability in terms of user revisit intention.*

Arunasalam Sambhanthan (*) researches on the application of human computer interaction techniques to key application areas such as business, education and health. He holds a Bachelors degree in Technology Management and Computing from the University of Portsmouth. His honors thesis investigated on an appropriate model for achieving strategic advantage in web tourism promotion with specific focus to Sri Lankan hotels as a case study for developing countries. Presently he is a researcher at the University of Portsmouth investigating on a second life based virtual therapeutic community for patients with Borderline Personality Disorder. Sam has published his research through international conferences and peer reviewed journals. He also serves as a reviewer for the Journal of Information, Information Technology and Organizations and Electronic Commerce Research and Applications. Recently he has been invited to serve as a member of the international board of reviewers of the 2012 InSITE conference organized by the Informing Science Institute, USA.

Dr Alice Good lectures in Human Computer Interaction and Strategic Business IT, as well as previously lecturing in e-commerce. She is also supervising a number of distance learning PhDs. Research expertise in accessibility, user centered design and mental health support systems. Other research interests include e-learning and e-commerce. Currently co-coordinating a multi-disciplinary project in providing virtual and mobile support for people with mental health problems.  Previous projects have looked at developing algorithms to rate the accessibility of web pages and the evaluation of learning environments.  Publications include refereed book chapters, journal and conference papers.  Invited reviewer for the: Interact conference in HCI, International Conference on Intelligent User Interfaces and the Transaction on Interactive Intelligent Systems Journal.

Arunasalam Sambhanthan
Department of Computing,
University of Portsmouth,
Buckingham Building, Lion Terrace,
Portsmouth PO1 3HE,
United Kingdom
arunsambhanthan@gmail.com

Alice Good
Department of Computing,
University of Portsmouth,
Buckingham Building, Lion Terrace,
Portsmouth PO1 3HE,
United Kingdom
alice.good@port.ac.uk